\documentclass[manuscript,screen,acmsmall]{acmart}

\usepackage{siunitx}

\usepackage{longtable}
\usepackage{lipsum} 

\AtBeginDocument{%
  \providecommand\BibTeX{{%
    \normalfont B\kern-0.5em{\scshape i\kern-0.25em b}\kern-0.8em\TeX}}}

\setcopyright{acmlicensed}
\acmJournal{TSC}
\acmYear{2022} 
\acmPrice{15.00}\acmDOI{10.1145/3532102}
\begin{document}

\title{Modeling Political Activism around Gun Debate via Social Media}

\author{Yelena Mejova}
\email{yelenamejova@acm.org}
\orcid{0000-0001-5560-4109}
\affiliation{%
  \institution{ISI Foundation}
  \city{Turin}
  \country{Italy}
}

\author{Jisun An}
\orcid{0000-0002-4353-8009}
\affiliation{%
  \institution{Singapore Management University}
  \country{Singapore}
}

\author{Gianmarco De Francisci Morales}
\orcid{0000-0002-2415-494X}
\affiliation{%
  \institution{ISI Foundation}
  \city{Turin}
  \country{Italy}
}

\author{Haewoon Kwak}
\orcid{0000-0003-1418-0834}
\affiliation{%
  \institution{Singapore Management University}
  \country{Singapore}
}

\renewcommand{\shortauthors}{Mejova et al.}

\begin{abstract}
The United States have some of the highest rates of gun violence among developed countries.
Yet, there is a disagreement about the extent to which firearms should be regulated.
In this study, we employ social media signals to examine the predictors of offline political activism, at both population and individual level.
We show that it is possible to classify the stance of users on the gun issue, especially accurately when network information is available. 
Alongside socioeconomic variables, network information such as the relative size of the two sides of the debate is also predictive of state-level gun policy.
On individual level, we build a statistical model using network, content, and psycho-linguistic features that predicts real-life political action, and explore the most predictive linguistic features. 
Thus, we argue that, alongside demographics and socioeconomic indicators, social media provides useful signals in the holistic modeling of political engagement around the gun debate.


\end{abstract}



\maketitle

\section{Introduction}
\label{sec:intro}

The issue of gun control vs. gun rights is a long-standing controversy in the United States.\footnote{\url{http://www.nytimes.com/interactive/2017/10/05/upshot/gun-ownership-partisan-divide.html}}
On the one hand, the U.S. has one of the highest rates of gun-related deaths among developed OECD countries~\cite{grinshteyn2016violent}.
On the other hand, the right to own guns is constitutionally protected by the United States Bill of Rights (the first ten amendments of the U.S. constitution).
The picture is, however, heterogeneous across the different states; states with more permissive gun laws and greater gun ownership have higher rates of mass shootings~\citep{Reepingl542}.
There have been numerous attempts at understanding the reason behind this deep ideological divide in the population~\cite{kalesan2016gun,chemerinsky2004putting}.
Gun control advocates typically blame gun violence on the pervasive presence and availability of guns.
Conversely, gun rights supporters usually blame gun violence on other cultural issues such as the portrayal of violence in media, lack of family values, or mental health issues~\cite{stone2015mass}.
Clearly, the issue of gun control vs. gun rights is a political and ideological one.
This complex issue has been studied in various contexts, such as political~\cite{chemerinsky2004putting}, cultural~\cite{kalesan2016gun}, and news media~\cite{stone2017boy}. 
The intense debate has moved onto online--gun control has been considered as one of the most controversial issues presented social media and both views have been reinforced and propagated via social networks \cite{garimella2018quantifying}. 
Since Twitter has emerged as a model organism for studying controversial issues and political discourse~\citep{conover2011political,garimella2018political,grinberg2019fake,barbera2015understanding,romero2011differences,mejova2015twitter}, and forecasting offline activism~\cite{ertugrul2019activism}, it provides a great opportunity to study and better understand the gun control political activities. 


In this study, we explore the factors related to this ideological divide through the lens of digital debate and offline political action.
We combine existing political science theories, the latest  computational tools, and a large dataset of digital traces, to systematically study online social responses to mass shootings.
Our aim is to examine simultaneously multiple aspects of political mobilization online, including the importance of network structure, content diversity, and emotion, alongside the standard demographic variables.
Thus, we add online information to a \textbf{holistic picture of the factors associated with political engagement in the gun debate}. 

To this end, we create a large Twitter dataset which encompasses the discussion in the aftermath of the deadliest mass shooting by an individual in U.S. history thus far---the 2017 Las Vegas shooting\footnote{\url{https://en.wikipedia.org/wiki/2017_Las_Vegas_shooting}} where 59 people were killed and 412 wounded. 
As a result of this mass shooting, the U.S. justice department imposed new regulations that banned bump stocks, which allow to fire a semi-automatic weapon in rapid succession, thus mimicking an automatic one.
A few months later, another mass shooting in Parkland, Florida\footnote{\url{https://en.wikipedia.org/wiki/Stoneman_Douglas_High_School_shooting}} spurred the organization of the so-called ``March for our lives'',\footnote{\url{https://en.wikipedia.org/wiki/March_for_Our_Lives}} a nation-wide protest to ask for stricter gun control laws.
This data captures the political mobilization these events have precipitated, thus making an examination its precedents possible.


Guided by previous work on sociology and psychology of political engagement, we design content, behavioral, and network features that can be extracted from Twitter.
For instance, guided by Granovetter's ``embeddedness'' argument~\cite{granovetter1985economic}, we compute various network features measuring clustering, assortativity, and connectivity, among others, for both groups and individuals.
Additionally, following the recent work around ideological ``echo chambers''~\cite{conover2011political,garimella2018political}, we capture the diversity of the discourse via network features and entropy of content elements such as hashtags, retweets, and unique words.
Finally, we enrich this data by using geo-location coupled with socioeconomic census data and state-level voting statistics.


Using these features, we model outcomes at both collective (state) and individual level.
First, we show that features extracted from Twitter are significant when modeling the gun policy of a state, even when controlling for other sociodemographic variables.
Second, we illustrate that it is possible to \emph{predict} individual attendance at the March for Our Lives based on public self-declaration, largely by using the content of their tweets before the event.

 
By introspecting both models, we are able to confirm the value of the signals coming from Twitter.
For instance, in the state model, some of the most predictive variables are based on the Twitter network\textemdash remarkably, with a larger effect than the voting preferences of the state.
Meanwhile, at the individual level tweet content features related to politics, emotion, and self-reference outperform demographic ones. 
Thus, we illustrate the usefulness of Twitter signals in modeling the online~\textemdash~and especially \emph{offline}~\textemdash~political engagement, and thus contribute to the ongoing development of theory-driven big data methods for computational social science.

Crucially, beyond sociopolitical insights, this work brings critical \textbf{ethical implications} by revealing a risk for grassroots movements.
All data collected in this study is publicly available, and no individual-level demographic information was used. 
Yet, on a filtered \emph{class-balanced} dataset, we are able to predict whether an individual attends the March for Our Lives and makes their attendance known at F1 = \num{82}\% with off-the-shelf tools. 
While these users are a subset of those who attended the protest, the risk that predictive tools create is underestimated by many~\cite{hinds2020wouldn}.
It is particularly significant in countries where democratic rights are not well-protected, and deserves to be better studied and understood.
Thus, it is imperative to raise awareness of this vulnerability, which may drive the design of both digital platforms, and the political campaigns that use them.

\section{Related Work}

Social media has been acknowledged as an important part of mobilization and communication during political movements~\cite{anderson2018activism}. 
Below we outline some major directions in computational social science research of political organization on social media and provide related theories of the social and psychological mechanisms involved.

\subsection{Social Media and Political Movements}

Internet use has been linked to political involvement through several channels.
First, social media may support civic education via strengthening existing ties and conveying political information through trusted networks~\cite{tang2013facebook,xenos2014great}.
Second, it may increase the chance of encountering information about a political organization~\cite{lotan2011arab}.
Third, participation in political discussion may develop an interest in further involvement in political discourse~\cite{halpern2013social}, as long as the experience of political deliberation on social media is positive and does not trigger conflict avoidance~\cite{vraga2015individual}.
However, engagement on social media does not always translate into political activities requiring a greater commitment of resources and may be confined to low-resource engagement such as viewing debates and posting comments (so-called ``slacktivism'')~\cite{vitak2011s}.

From the early days of social media, blogging has created an outlet outside of mainstream media to bring new issues into the public debate and set alternative framings to ongoing debates. 
Despite their small readership, because media elites (editors, journalists, publishers) eventually began consuming political blog content in the early 2000s, online discourse started to have a broader impact on politics~\cite{drezner2004power}.
The ensuing increased popularity of social media, as well as shortening of the posts \textemdash exemplified by Twitter's 140 character limit\textemdash has allowed for increased democratization of political expression. 
Although early attempts at predicting political events such as elections have been met with skepticism in the research community~\cite{Metaxas2011,gayo2012no,jungherr2012pirate}, a series of anti-government protests known as the Arab Spring, 
spurred new interest in the use of social media for political organization; some even coined the disputed shorthand ``Twitter Revolution'' for the early demonstrations in Tunisia, Libya, Egypt, and Syria~\cite{Lever2013}.

A survey of media use by Egyptian protesters in early 2011 shows that more than a quarter of respondents first heard of the protests on Facebook, and a quarter used Facebook to disseminate pictures and videos they produced~\cite{tufekci2012social}. 
Conversely, those who used blogs and Twitter for both general information and for communicating about the protests were more likely to attend on the first day of the protests, though nearly half of the respondents first heard about the protests from someone face-to-face~\cite{tufekci2012social}.
However, a content analysis found that, out of Twitter users participating in the discussions in Tunisia and Egypt, more than a quarter were institutional accounts, and among individuals, journalists and activists were the most common sources of information~\cite{lotan2011arab}, thus putting in question how many ordinary citizens took an active role. 
The repercussions of these movements continue to echo in the politics of the region and its expression on social media, with the accompanying research on the rise of ISIS~\cite{magdy2016failedrevolutions} and government messaging~\cite{darwish2017seminar}.

Protest movements elsewhere in the world have also been captured on social media, including Spanish protests in May 2011~\cite{gonzalez2011dynamics}, Occupy Wall Street in the fall of 2011-2012~\cite{conover2013digital,conover2013geospatial}, and Black Lives Matter in 2014-2015~\cite{freelon2018quantifying}. 
The latter work examines the power dynamic of ordinary citizens and the political elites and shows a stronger Granger causality with the activity by elite accounts following that by ordinary posters~\cite{freelon2018quantifying}. 
Despite the potential of global involvement, such movements tend to be highly localized, such as the Occupy Wall Street, during which users in New York, California, and Washington D.C. produced more than half of all retweeted content, and with content at local scale focusing on locations and timings of the events, while cross-state communication focusing on the core framing of the issue~\cite{conover2013geospatial}.  
Beyond political movements, experimental studies show that social media messaging is effective in spurring political action, such as in a large Facebook study that shows informational messaging increases voter turnout for users who saw the message and for their friends~\cite{bond201261}.
Survey-based studies often find supporting evidence of the relationship between social media use and political activism.
A study of Chilean protests found that the effect of social media on the likelihood of political engagement is comparable to the influence of education and participation in civic groups~\cite{valenzuela2013unpacking}. 
A meta-analysis of such surveys in 2015 has found more than $80\%$ of coefficients to be positive~\cite{boulianne2015social}, but these coefficients are slightly more likely to be statistically significant in well-established democracies, and in samples of youths.
Thus, precisely which political actions, where, and by whom are more likely to be influenced by social media is an ongoing topic of debate.

\subsection{Determinants of Political Engagement}

Social media is, by far, not the only force behind political engagement.
Indeed, strong demographic, socioeconomic, and ideological forces drive political activity, some of which we briefly outline below (for a more extensive discussion we refer the reader to political science literature~\cite{opp2009theories}).

\subsubsection{Demographics}

In the previous years, the demographics of U.S. electorate has been changing markedly. 
The U.S. election in 2018 saw a marked increase in the rates of voters of racial minority groups, who were more likely to cast their ballots early by mail~\cite{krogstad2019historic}. 
Yet, there are still findings that strict identification laws suppress the turnout of black, hispanic, asian, and multi-racial groups~\cite{hajnal2017voter} (although the quality of survey data has been criticized in this domain \cite{grimmer2018obstacles}). 
In the same year, for the first time Generation Z, Millennials, and Generation X voters outnumbered Baby Boomer and previous generations~\cite{cilluffo2019genz}.  
Ongoing changes in the socioeconomic status is also seen as a complex contributor to the participation in both in radical left and right parties~\cite{rooduijn2018paradox}, and in populist movements~\cite{van2019populist}.
The polarization among demographic lines is expanding, with race, gender, and education being increasingly associated with diverging political leanings~\cite{tyson2018midterm}.
In this work, we attempt to capture as many demographic and socioeconomic attributes of the social media users as possible in order to provide a strong baseline for modeling gun debate engagement.

\subsubsection{Social Network}

In ``Social Structure and Citizenship''~\citet{scheufele2004social} outline a framework of political engagement analysis that connects macroscopic sociological variables and individual-level behaviors, and illustrate via a national survey that a ``social setting in which citizens discuss politics is an important antecedent of political participation.''
Such research juxtaposes the traditional individual-level focus on ``individual's party identification, issue positions, ideology, and images of candidates''~\cite{campbell1980american}, with the new works inspired by Granoveter's  ``embeddedness'' argument, which stresses ``the role of concrete personal relations and structures (or networks) of such relations in generating trust and discouraging malfeasance''~\cite{granovetter1985economic}.
An even earlier study of discussion networks~\cite{leighley1990social} shows increasing levels of discussion diversity (network heterogeneity) is linked to an increase in traditional forms of political participation.
In particular, discussant conflict, to which degree the person is exposed to the conflicting opinions, is positively and strongly associated with voting and campaign activities. 
However, the size of a person's network of political discussants is negatively related to voting, but positively to campaign activities.

Social media has provided an alternative avenue for analyzing social ties.  
For instance,~\citet{valenzuela2018ties} hypothesize that different social media platforms influence political participation differently. 
Their cross-sectional, face-to-face survey of a representative sample of Chilean youths shows that different platforms represent different types of social ties that may result in political activity, with Facebook friendships better capturing ``strong'' ties and Twitter followership ``weak'' ties~\cite{granovetter1983strength}.
Moreover, social network use is associated with civic and political engagement, but also with the increased perception of social capital, even after taking into account demographics and news media use~\cite{gil2012social}.
In yet another survey, respondents with larger Facebook networks also participated in politics more frequently, though the impact of network size was completely mediated by network structural heterogeneity and connection with public political actors~\cite{tang2013facebook}.
Online response to major events are also an opportunity to model political opinion formation.
\citet{an2016you} examine the response to the 2015 Charlie Hebdo newspaper shooting in the light of three theories: \emph{Clash of Civilizations} that predicts social fault lines between major ``civilizations''~\cite{huntington1993clash}, \emph{Density Theory} that puts importance on the individual's social context~\cite{huckfeldt2009interdependence}, and \emph{Interdependence Theory} that further emphasizes the importance inter-personal relationships~\cite{thibaut1959social}.
Combining information on user location, followership, and mention, they show a stronger support for the latter two theories.

Our models also include network features, such as size and heterogeneity, as well as user-specific features of centrality.

\subsubsection{Ideology}

Ideally, social media may be envisioned as a forum for ``deliberation''~\cite{habermas1991structural} wherein there is a ``interchange of rational-critical arguments among a group of individuals, triggered by a common or public problem, whose main focus or topic of discussion is to find a solution acceptable to all who have a stake in the issue''~\cite{halpern2013social}. 
However, even before social media, political campaigns have been seen as domains of ``low information rationality'' where ``individuals rely on ideological heuristics or similar shortcuts instead of carefully collecting and selecting all available information''~\cite{scheufele2004social}.
Cognitive biases often prevent people from accurately perceiving information that contradicts their existing opinion.
For instance,~\cite{maccoun2009citizens} found that citizens, ``especially those holding conservative beliefs, tended to attribute studies with liberal findings to the liberalism of the researcher, but citizens were less likely to attribute conservative findings to the conservatism of the researcher''.
The resulting ``echo chambers'' are often seen in the structure of the communication networks on social media~\cite{conover2011political,garimella2018political,bakshy2015exposure,barbera2015tweeting}.
Inside these closed communities, the information spread is homogeneous and often biased, thus fostering homophilic attitudes~\cite{cota2019quantifying,garimella2017effect}.
Indeed, it has been shown that people prefer media that conforms with their attitudes in their daily media diet \cite{johnson2009communication}, although not necessarily to the exclusion of opposing sources \cite{stroud2008media}.
On social media, various factors contribute to the users' content preferences, including endorsements from their social network \cite{winter2016selective} and personal relevance \cite{mummolo2016news}.
To what degree the selective content consumption on social media is attributable to psychological or social tendencies, and what role recommender and search algorithms play in the potential reinforcement of these tendencies is difficult to measure, as the control over, and knowledge of, these algorithms are proprietary \cite{guess2018avoiding}.
We must note, however, that the communities displaying content and communication polarization which have been documented on Twitter represent a particularly partisan portion of the overall media consumption, since most of the news consumption takes place from the large mainstream sources \cite{nelson2017myth}. 
Still, a 2021 Pew Research Center survey found that ``[a] little under half (48\%) of U.S. adults say they get news from social media `often' or `sometimes''' with Twitter being the second most popular website after Facebook \cite{walker2021news}. 
Monitoring whether the growing use of social media contributes to polarization on important societal issues is imperative to the understanding the deliberation process related to eventual policy decisions.

Ideological polarization is especially pronounced in the U.S. gun control debate \textemdash a 2016 Pew Research Center poll revealed that just 22 percent of self-identified Democrats chose protecting gun ownership rights, compared with the 76 percent of Republicans who did \textemdash a gap which has widened since 2007 mainly by Republicans adopting a stronger stand on the issue~\cite{enten2017guns}.
In fact, although the question of gun legislation has not been a salient political issue until the 1960s~\cite{crocker1982attitudes}, in the 2000s it has become one of the most divisive issues between the U.S. Democrat and Republican parties~\cite{kleck2017point}.
Beyond the political partisanship, personal inclinations towards individualism as opposed to collectivism are associated with peoples' attitudes toward gun ownership and gun control~\cite{celinska2007individualism}.
To which extent this hyper-polarization is revealed in social media debates, and how it is expressed via local gun regulation is an exciting research question.
In this work, we attempt to include both political and gun-specific attitudes, as well as more general diversity in language, that may be related to political opinion and action.

\subsection{Privacy concerns of tracking political movements}

In early days, research on political movements on Twitter can alleviate the information asymmetries between protesters and authorities, such as protest locations, police locations, and police actions, which had been seen as `monopolized' by the authorities~\cite{earl2013this}. 
The danger of social media surveillance by authorities is also noted by \citet{owen2017monitoring} with several attempts to monitor protesters' social media activity around Occupy protest. 
Since the Cambridge Analytica scandal~\cite{hinds2020wouldn}, the awareness of social media-based surveillance has also been permeating the public. 
According to the report by Pew Research in 2014~\cite{madden2014public}, most of the U.S. adults have heard about ``the government collecting information about telephone calls, emails, and other online communications as part of efforts to monitor terrorist activity.'' 
The awareness of government monitoring changes many internet users' behaviors and practices by themselves~\cite{barnard2012surveillance}. 
Yet, many people share their offline political activities (e.g., attending a protest) on social media.
The present study attempts to predict whether an individual attends the March for Our Lives (and makes their attendance known) by using public Twitter data only. Our results thus warn of the potential risk that those users can be identified with high accuracy, even when 
they do not explicitly declare their intentions ahead of time.

\section{Methods}
\label{sec:methods}

Below we describe the steps taken for collecting the data from Twitter, user filtering, and stance classification using network and content information.
We then describe the multifaceted modeling of the two sides of the debate, as well as the individuals within, using demographics extracted from the U.S. Census, Twitter attributes, network features, and linguistic features including sentiment and hate speech.
In particular, we consider two tasks: (1) at the state level, modeling strictness of the state's gun laws, and (2) at the user level, predicting public disclosure of participation to the ``March for our Lives'',\footnote{\url{https://marchforourlives.com}} a student-led demonstration in support of gun control laws.

\subsection{Data}

We collect tweet data using Twitter's Streaming API from Oct 1, 2017 to May 1, 2018 using the keywords \emph{gun}, \emph{guns}, and \emph{nra} (National Rifle Association).
The collection resulted in capturing \num{142 874 864} tweets written by \num{13 809 004} users.
This time span includes the social responses  after the Las Vegas mass shooting which happened on October 1, 2017 and the subsequent March for our Lives held on March 24, 2018.

\subsection{User Selection}
\label{sec:userselection}

As Twitter is notorious for having automated accounts (bots)~\cite{davis2016botornot}, we apply a number of filters to ensure that the accounts to be examined are not automated, and have enough activity to be modeled.
Additional filters were applied to make sure the geo-location of the users was reliable, by ensuring the user has tweeted at least twice from the same state.
The filters below result in \num{1 857 749} users, who posted \num{46 943 763} tweets.

\begin{itemize}
\setlength\itemsep{-0.2em}
\item Exclude users with one tweet in our dataset.
\item Exclude top 0.1\% of users by number of tweets in our dataset.
\item Exclude users with fewer than 5 followers and 5 friends.
\item Exclude users whose friends to followers ratio is more than 10 (inspired by \cite{wang2010detecting}).
\item Exclude users whose account is less than 1 year old.
\item Exclude those not having at least 1 English tweet in our dataset.
\item Exclude users whose location cannot be mapped to one of U.S. states (or DC).
\item Exclude users who do not have more than one tweet from the same state.
\end{itemize}

Geo-location is performed firstly by using the GPS coordinates embedded in a tweet (a small minority of tweets have such metadata), and secondly, by using Yahoo! Placemaker API\footnote{\url{https://www.programmableweb.com/api/yahoo-placemaker}} applied to the \texttt{location} field of the user profile. 
The returned geo-mappings by the geo-location service were manually checked for the \num{1000} most popular locations in the dataset to ensure the correct outcome for most of the users.

\subsection{Stance Classification}

In order to classify the stance of the captured users on the gun issue (as either pro gun rights or pro gun control), we employ two sources of information: (1) the retweet action, as it usually indicates agreement with another user, and (2) the content of the user's tweets, as it conveys the actual stance of the user. 

Following previous work on controversy on social media~\cite{garimella2016quantifying}, we build an \emph{endorsement graph} in which users are mapped into nodes and retweets are mapped into (directed) edges, weighted by the number of times one user has retweeted another. 
We apply a threshold of 2 on the edges to reduce noise, or in other words, we eliminate one-time retweet relationship from the graph. 
We then use the graph partitioning algorithm METIS~\cite{karypis1998fast} to partition the network into two groups.
We partition the graph repeatedly $N$=100 times with different random seeds to get an ensemble of partition assignments for each node, and use the average partition assignment for each node (i.e. user) across the $N$ repetitions as a polarity score $p \in [0,1]$~\citep{cossard2020falling}.\footnote{We optimize the parameter indicating the relative size of the two sides by maximizing the number of users within 95\% confidence interval of either extreme, finding the optimal proportion to be 1.5 to 1, with gun control being the larger side.}
The partitioning obtained by METIS with 2 partitions has a modularity of $0.427$.
As a comparison, partitioning the same graph with the Louvain algorithm~\citep{blondel2008fast} achieves a modularity of $0.480$ with a number of partitions larger than 100k.
The small difference between the two modularity scores confirms that this approach captures the main structure within the graph.
As the algorithm does not indicate the actual stances of the users (in fact, it uses no tweet content information whatsoever), each partition is assigned its leaning manually, by examining 10 sampled users on each side.
In other words, the leaning of the sampled users is then ``propagated'' to all users within their partition.

As the graph partitioning algorithm partitions the giant connected component (GCC) only, for those users who are not in the GCC of the retweet network, we train a content-based classifier on users who have been classified.
After under-sampling the majority class (pro gun control) to balance the dataset, the vocabulary is constructed by removing stop words and punctuation.
We test three classifiers, including Support Vector Machines, Naive Bayes, Logistic Regression, and Random Forest, by using 5-fold cross-validation. 
Naive Bayes using frequency-based vectorizer performs the best, and achieves 99.4\% accuracy during cross-validation on users classified by using the retweet network.
We then apply this classifier to the unlabeled users ---those not in GCC of the retweet network--- with the class membership determined by using 1\% error threshold (users with a probability greater than 0.99 or smaller than 0.01 are assigned with the label).

\subsection{Model 1: State Gun Laws}

Our first model addresses the question---\emph{how does the Twitter gun debate relate to the gun control political activity, as measured via gun laws passed in each state?}
We define the target variable as the strictness of gun laws in a state, encoded as a rating from $1$ (most strict) to $5$ (most friendly) extracted from \emph{Guns To Carry},\footnote{\url{https://www.gunstocarry.com/gun-laws-state}} a resource that summarizes the laws around the country, which we have also verified with additional information on each state's laws.\footnote{\url{https://statelaws.findlaw.com/criminal-laws/gun-control.html}}
For example, in 2019 California was one of the most restrictive states, requiring permits both for handguns and long guns, and allowing open carry only in some counties (it has a rating of $1$). 
Kansas, on the other hand, requires no permits or registration, and allows open carry, and has a rating of $5$.

The independent variables are meant to capture the peculiarities of the gun debate on Twitter for a particular state. 
In order to tease out the relationship between Twitter and state laws, we employ several controls, including demographic composition of the state, socio-economic and wellbeing indicators, gun culture (as measured by gun sales), and political inclination of the population. 
The features, computed for each state, are listed in Table \ref{tab:statefeatures}.
All the features have been standardized and transformed to make them approximately normal, so that the coefficients of the model can be compared.
Demographic and socio-economic information come from 2018 County Health Rankings,\footnote{\url{https://www.countyhealthrankings.org}} a resource which combines official census and health statistics from U.S. governmental agencies. 
The gun sales variable is obtained as a proxy from the FBI National Instant Criminal Background Check (NICS) database.\footnote{\url{https://www.fbi.gov/services/cjis/nics}}
Political inclination is measured as the percentage of the state's population that has voted for a Republican candidate in the 2016 U.S. Presidential Election.\footnote{\url{https://ballotpedia.org/Voter_turnout_in_United_States_elections}}
The Twitter Network variables are computed both for all users within the state, and for the communities on the two sides of the debate. 
The number of nodes, edges, clustering coefficient, and density are also computed on the out-edge induced subgraph in order to capture interactions with entities outside the given state.
Finally, Twitter Content and Account features are computed as averages over the users in the state, as well as for the two communities separately.
The hate dictionary comes from Hate Base project\footnote{\url{https://hatebase.org}} and sentiment lexicon from~\citet{10.1371/journal.pone.0026752}.
We also compute the average number of retweets, hashtags, and unique words used (vocabulary), as well as the entropy of the distribution of usage (which measures, intuitively, the diversity in behavior of the users). 
Twitter Account variables include average user follower and friend (``followee'') counts, as well as daily tweet rate when considering all tweets, those about guns, and those in English.

As there are only 51 data points (50 states and Washington DC) in the dataset, we apply several steps to decrease the number of variables.
We begin by computing univariate correlations between the features and the target variable, and select only variables having Pearson correlation with a magnitude of at least $0.3$. 
We then perform a Variance Inflation Factor (VIF) multicollinearity check, and remove variables with the highest VIF until all variables have a VIF of under $6$. 
The resulting 18 variables are then used to build a Structural Equation Model (SEM)~\cite{hoyle1995structural,hox1998introduction} that relates each of these variable groups to the outcome variable of state gun law rating, described above. 
We choose SEM instead of a simple linear regression because the variable groupings alleviate some of the multicollinearity of the data.
We use the Latent Variable Analysis (\texttt{lavaan}) R library\footnote{\url{http://lavaan.ugent.be}} to fit the parameters of the model.


\begin{table}[t]
\centering
\caption{Features of state-level gun discussion. Network features with ``+EIG'' were also computed on out edge induced subgraph.}
\label{tab:statefeatures}
\begin{tabular}{p{0.35\textwidth} p{0.35\textwidth}}
  \toprule
\textbf{Demographics} & \textbf{Twitter Network} \\
perc under 18 & $|$nodes$|$ (+EIG), $|$edges$|$ (+EIG) \\
perc 65 and over & clustering coefficient (+EIG) \\
perc African American & density (+EIG) \\
perc Asian & max edge weight \\
perc Hispanic & degree assortativity \\
perc non Hispanic white & Gini coefficient of in-degree \\
perc rural & node proportion deviation \\
 &  \\
\textbf{Socio-Economic} & \textbf{Twitter Content} \\
high school graduation rate & retweet count \\
perc some college & retweet entropy \\
perc unemployment & hashtag count \\
income inequality ratio & hashtag entropy \\
perc uninsured & vocabulary count \\
perc single parent households & vocabulary entropy \\
perc association rate & hate word rate \\
violent crime rate & avg sentiment polarity \\
perc severe housing problems &  \\
median household income & \textbf{Twitter Behavior} \\
residential segregation black white & user followers count \\
homicide rate & user friends count \\
 & follower/friend ratio \\
\textbf{Health} & account age \\
mentally unhealthy days & gun tweet rate \\
perc adult smoking & allt weet rate \\
perc adult obesity & gun tweet count \\
perc excessive drinking & English tweet count \\
 &  \\
\textbf{Politics \& Gun Culture} & \textbf{Target} \\
gun sales & gun law rating \\
firearm fatalities rate &  \\
perc vote republican &  \\
\bottomrule
\end{tabular}
\end{table}

\subsection{Model 2: Individual March Attendance}

Our second model considers the individual -- \emph{is it possible to predict individuals who will publicly engage in offline political protests 
by using their prior online activity?}
We operationalize political engagement in offline protests by considering individuals who have left tweets to indicate that they have attended one of the ``March for Our Lives'' events held across the United States on March 24, 2018.\footnote{\url{https://en.wikipedia.org/wiki/List_of_March_for_Our_Lives_locations}}
As mentioned in Section~\ref{sec:intro}, this experiment also aims to call attention to potential risks that protesters, even those who do not publicly declare their attendance, can be identified by prior online activities, a risk still underestimated by many~\cite{hinds2020wouldn}.

Specifically, we select tweets (original, not retweets) which have geo-location or match the regular expressions ``I am going to'' or ``I am at'', posted on the day of the march.
To make sure the tweets are of people attending the march, we employ the CrowdFlower crowdsourcing platform (now called Appen) to label whether the user has indeed attended the march, or just tweeted in support or opposition.
The task is fairly easy, with annotator agreement (measured via label overlap) high at 92\%; 310 labelers participated in the task, with final label determined via majority rule over 3 labels from different annotators, for each tweet.
The resulting \num{1343} unique users are identified as having declared attending a march, although only \num{658} pass the user filtering steps described in Section~\ref{sec:userselection}. 
Out of these, vast majority were identified as pro gun control (99\%) and only 7 as pro gun rights.
To provide an even comparison of these users to those who were not detected to have attended the march, we perform a random sampling of the rest of the users, stratified by state in proportion to the original distribution of the dataset, which results in a control set of 610 users. 

The features, computed on the tweets posted \emph{prior} to the march, largely overlap with those listed in Table~\ref{tab:statefeatures}. 
The demographic features are taken in accordance to the county in which the user has been detected to have tweeted the most.
The Twitter network features, however, change as they concern an individual node---these are listed in Table~\ref{tab:usernetworkfeatures}.
Additionally, we include the lexical categories of Linguistic Inquiry and Word Count (LIWC),\footnote{\url{http://liwc.wpengine.com}} a listing of keywords with associated psychologically meaningful categories, 73 in total, including those concerning general grammar use (pronouns, articles, prepositions), emotion (positive, negative), and topics including body, feeling, work, and leisure.
While there are several efforts to improve LIWC by using modern NLP techniques~\cite{fast2016empath}, we use LIWC because its broad applicability has been proven across a wide range of domains~\cite{tumasjan2010predicting,abbar2015you,an2019political}.
We remove features that do not have any value for more than half of the users, 
and then removed rows with any missing values, resulting in a dataset of \num{1206} users (\num{631} of which attended the March For Our Lives).
The full list of features can be found in the Appendix~\ref{app:features}.
We use three widely-used models for prediction: Support Vector Machine (SVM) with linear kernel, Logistic Regression (LR), and Random Forest (RF). 
For SVM, we set the parameter values as $C$=1.0, kernel=`linear', gamma=`scale', class\_weight=`balanced.' For LR, we set the parameter values as $C$=1.0, penalty=`l2', class\_weight=`balanced.' For RF, we set the parameter values as n\_estimators=10, criterion=`gini,' min\_samples\_split=2, class\_weight=`balanced.'
Since the two classes are slightly imbalanced, we re-weight the classes inversely proportional to their frequency: $w_j = \frac{n}{kn_j}$, where $w_j$ is the weight to class $j$, $n$ is the number of observations, $n_j$ is the number of observations in class $j$, and $k$ is the total number of classes, which is 2 in our case.
The precision, recall, and F1 measures are then computed via 5-fold cross-validation.

\begin{table}[t]
\centering
\caption{Network features for users, computed on Twitter retweet network.}
\label{tab:usernetworkfeatures}
\begin{tabular}{p{0.35\textwidth} p{0.35\textwidth}}
  \toprule
indicator whether in GCC &  personalized PageRank \\
 in-degree & clustering coefficient \\
 out-degree & avg in-neigh in-degree \\
 max in-edge weight & avg out-neigh in-degree \\
 max out-edge weight & \\
\bottomrule
\end{tabular}
\end{table}

\subsection{Data Availability}
Due to Twitter Terms of Service we cannot publish the entire dataset (as tweets which may have been removed by their poster may still be in the data).
Instead, the terms allow to share a list of the numeric IDs of tweets which can be re-collected.
We will make it available to the research community upon request. 
In addition, we will make public all labeled data used for validation of the classifiers.


\section{Results}
\label{sec:results}

\subsection{User Classification}

\begin{figure}[t]
\centering
\includegraphics[width=0.80\linewidth]{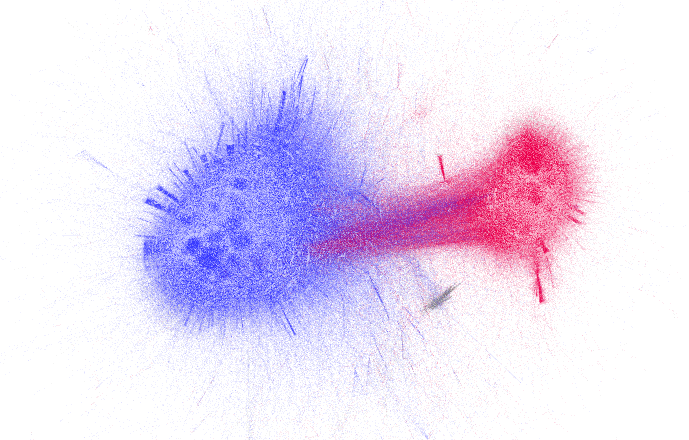}
\caption{Giant connected component of endorsement (retweet) network. METIS partitions are colored in red for pro gun rights and  blue for pro gun control.}
\label{fig:retweetnetwork}
\end{figure}

The endorsement (retweet) network and its partitioning by METIS~\cite{karypis1998fast} are shown in Figure \ref{fig:retweetnetwork}, wherein a force-directed layout is used to position the nodes. 
Out of the \num{125997} users in the giant connected component (GCC) of the retweet network, \num{82023} (65.1\%) are classified as pro gun control and \num{43974} (34.9\%) as pro gun rights. 
As visible in the figure, the two sides occupy distinct closed communities, with few links between the partitions. 
In order to evaluate the performance of this network-driven classification approach, four authors manually labeled 200 random users (100 from each side).
The classifier accuracy is 96\% on this manually-labeled test set. 
Annotator agreement, measured by label overlap, is extremely high at 98\%, thus indicating that users in the center of the retweet activity are fairly easy to classify.

The users labeled by using the network method are then used to train a content-based classifier, as described in Section~\ref{sec:methods}. 
We apply the classifier to the \num{1738759} users who were not included in the GCC of the retweet network, and find a total of \num{1274212} ($73.28\%$) pro gun control and \num{288793} ($16.61\%$) pro gun rights users. 
The remaining $10.11\%$ who do not pass the threshold are labeled as ``unknown'' and excluded from the further analysis.   
To evaluate the performance of this content-based classifier, the authors manually label 200 random users into three groups: pro gun control, pro gun rights, and unknown inter-annotator agreement for this task is high, with a Cohen's kappa of $0.87$).
Our content-based classifier achieves an overall accuracy of $70.0\%$.
Compared to a random baseline, which would achieve $52.6\%$ accuracy, this result makes the content-based classifier $36.7\%$ closer to a perfect classifier (i.e., its kappa statistic is $0.367$~\citep{bifet2015efficient}).
If we consider our content-based classifier as a quantification tool to understand the proportion of different classes in the dataset, this proportion is very close to the ``true'' proportion (taken from the manual labels). 
Performing $\chi^2$ test, we cannot reject the null hypothesis that the distributions are the same ($p=0.437$), which implies that they are very close. 

As an exploratory analysis, we characterize pro gun control and pro gun rights groups by investigating unigrams that are over-represented in each group. 
To compute the over-representation, we use log-odds ratios with informative Dirichlet priors~\cite{monroe2008fightin} computed by using background word frequency on the entire dataset.
The 50 most over-represented words (whose frequency is larger than 100k) for pro gun control and pro gun rights users are presented in Table~\ref{tab:overrepresented_words}.
On the gun control side, the most prominent is a person: Shannon Watts (\@shannonrwatts) is a founder of MomsDemand (\@MomsDemand), a grassroots movement of Americans ``demanding reasonable solutions to address our nation's culture of gun violence'' (quoted from account description). 
Others are popular hashtags referring to the gun control movement (MarchForOurLives and NeverAgain), politics (GOP, congress, reform, Trump), and the shooting (violence, teacher, Sandy, Hook).
On the gun rights side, the most prominent is Broward, a county in Florida at the center of a local law debate on limiting the sale of large-capacity magazines,\footnote{\url{https://www.nytimes.com/2018/03/08/us/florida-gun-bill.html}} mentions of other political entities (Hillary, Israel, POTUS), media channels (NRATV, CNN), and mentions of the second amendment. 
Note that both sides mention political entities from the other side: GOP (Republican party) by gun control side, and (Hillary) Clinton from gun rights side, which indicates a pointed attention across party lines we do not see in retweet behavior.

\begin{table*}[t]
\centering
\caption{Lists of the top 50 representative words based on log-odds ratios from gun control and gun rights sides.}
\label{tab:overrepresented_words}
\begin{tabular}{p{0.47\textwidth} p{0.47\textwidth}}
  \toprule
  \textbf{Gun control} & \textbf{Gun rights} \\
  \midrule
  shannonrwatts, violence, gop, marchforourlives, prayers, krassenstein, students, white, thoughts, funder, congress, neveragain, reform, teacher, emma4change, trump, money, republicans, sandy, teachers, health, hook, today, douglas, edkrassen, children, momsdemand, survivors, cameron\_kasky, fucking, change, fuck, joyannreid, lobby, mikel\_jollett, stoneman, kylegriffin1, mental, enough, russian, sayshummingbird, igorvolsky, days, nowthisnews, sarahchad\_, speakerryan, longlivekcx, proudresister, action, stephenking & broward, defend, hillary, member, facts, nratv, hate, israel, cnn, owners, amendment, members, tomilahren, security, push, firearm, smoking, county, truth, knew, murder, government, potus, evil, advocates, constitution, blaming, murders, used, cruz, shooters, 2nd, didn, second, drugs, problem, themselves, firearms, away, attack, ever, protect, gt, california, shooter, freedom, carry, rate, or, membership \\
  \bottomrule
\end{tabular}
\end{table*}


\subsection{Model 1: State Gun Laws}

\begin{figure}[t]
\centering
\includegraphics[width=0.8\linewidth]{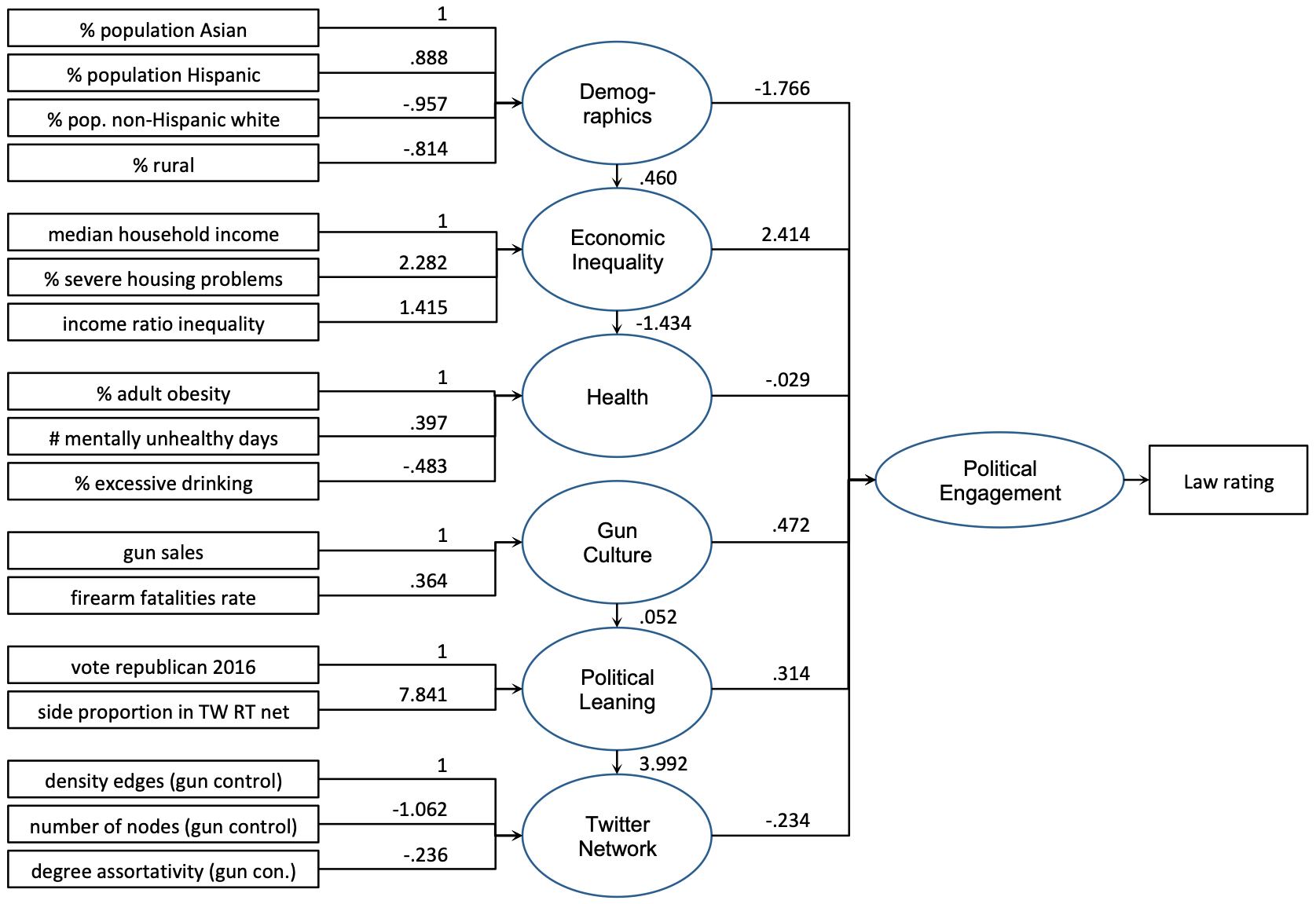}
\caption{Structural Equation Model of gun law rating (gun friendliness on a scale from 1 to 5). Coefficients are shown on the edges.}
\label{fig:sem}
\end{figure}

Next, we turn to modeling the state's gun laws by using characteristics of Twitter conversation about gun control, as well as demographic and socio-economic controls. 
The features spanning demographics, socio-economic, health, and culture, as well as Twitter-derived ones (listed in Table~\ref{tab:statefeatures}) are aggregated per state, and several feature selection steps have been applied to limit the number of variables.
We build a Structural Equation Model (SEM), with the variables used as indicators for six intermediate latent variables, as shown in Figure~\ref{fig:sem}.
Given the small number of items in the dataset (51 states), the fit statistics show the variables only partially explain the data, with Root Mean Square Error of Approximation (RMSEA) at $0.219$ (whereas commonly desirable values are below $0.06$~\cite{hu1999cutoff}).
We thus exclude the significance values in the figure and examine the coefficients only.
Intuitively, the effect strength of one variable on the outcome is a combination of all of the coefficients along all of the paths to the dependent variable. 
Based on related literature~\cite{opp2009theories,tyson2018midterm,leighley1990social,kleck2017point}, we also introduce effects between the latent variables, mainly linking demographics, economics, and health on one side, and gun culture, political culture, and Twitter network on the other.
Finally, despite the structure of the model, the resulting relationships should be thought of as correlational, and not causal.

First, we remind the reader that the higher gun law rating in a state, the fewer restrictions there are on gun ownership, and the more successful is the gun rights side. 
Demographics and social inequality latent variables show the strongest coefficients when relating to the law rating. In particular, percentages of Asian and Hispanic population have negative effect (more gun control), and percentages of non-Hispanic white and rural population have positive effect (more gun rights), with about same magnitude.
The strongest effect is by economic variables, especially percentage of severe housing problems (struggling with overcrowding, high costs, or lack of facilities) and income inequality, both associated positively with gun law rating (more gun rights). 
These effects are stronger than that of the median income, and much stronger than the demographic ones.
Health-related variables have the least effect on the gun laws, possibly because the demographic and economic dimensions of these variables are already accounted for (note the strong relationship between economic inequality and health).
On the cultural side, while we find gun sales and firearm fatalities to be positively associated with the gun law rating (more gun rights), these gun-related variables have a very slight effect on the political leaning latent variable.
Instead, we find a strong relationship between the political leaning and the Twitter network latent variables, and here also find the strongest effect in the model: proportion of nodes on gun rights side in Twitter retweet network, which is 7 times stronger than the proportion of population who voted for Republican.
Finally, the rest of the Twitter-related indicators show a negative relationship with gun law ratings.
We find that Twitter retweet networks of users identified to be on gun control side with higher edge density and lower number of nodes and lower assortativity (smaller communities with star-like structures) are associated with gun control laws.

In summary, we find socio-economic and demographic variables to be some of the most predictive of law ratings in a state.
However, we also discover a strong predictor from Twitter: the proportion of nodes in the two sides of the Twitter debate, with the higher proportion of gun rights to gun control, the higher the gun law rating of the state.

\begin{figure}[t]
\centering
\includegraphics[width=0.8\linewidth]{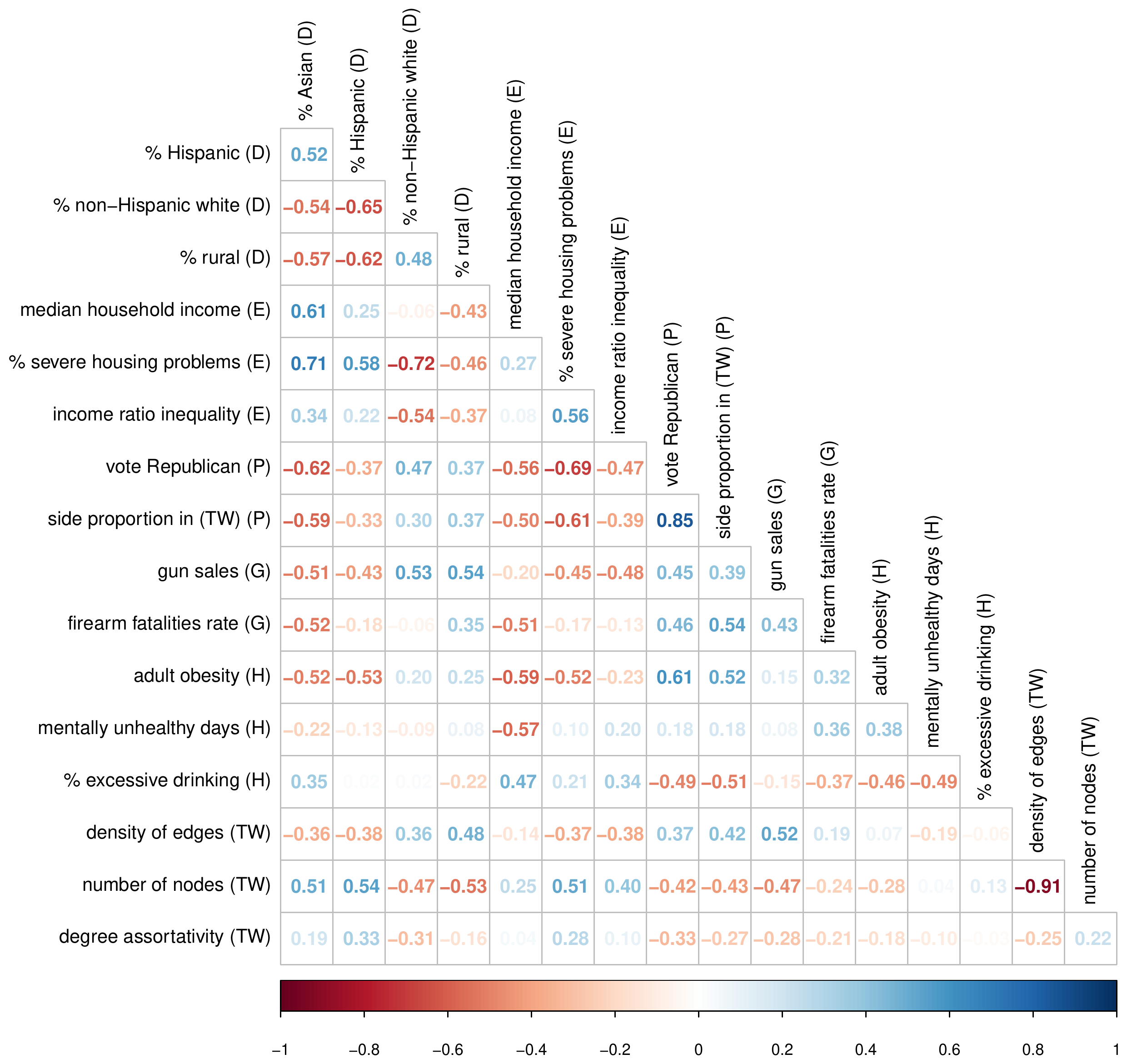}
\caption{Pearson correlations between the variables used in Figure \ref{fig:sem} and Table \ref{tab:ablation}.}
\label{fig:var_cor}
\end{figure}

Although we are constrained from making statistical significance claims due to the data size, we perform several checks in other to ascertain the importance of the features.
Table \ref{tab:ablation} shows the model fit measured using Multiple R$^2$ for a OLS regression models using subsets of features discussed above. 
The left column shows the fit of models using features of only one group, and right column that of models which use all features except those in a group.
The best single-group model contains the Demographic variables, followed by Gun Culture, and Economic Inequality. 
Whereas Twitter Network features provide the least information (similarly to Health features), when all Twitter features are considered (including those used to measure the political leaning of the state), the model achieves R$^2$ = \num{0.331}.
When considering the ablation experiments in the right column, we find that no exclusion of a single group of features brings the fit down substantially. 
Thus, we explore the relationship between the variables in Figure \ref{fig:var_cor} with a correlation matrix among the features.
Indeed, we find several striking relationships, including those between demographic and gun-related variables and Twitter-related ones.
For instance, the proportion of retweet network that is pro-gun rights is positively related to higher proportion of votes for Republican party ($r$ = \num{0.85}) and negatively to the median household income ($r$ = \num{-0.50}).
Furthermore, the proportion of rural areas in the state is positively related to the density of nodes in the retweet network ($r$ = \num{0.48}), and negatively to the number of nodes ($r$ = \num{-0.53}).
Thus, it is possible that the retweet network captures some of the relevant demographic and gun-culture specific attributes of the states.

\begin{table*}[t]
\centering
\caption{Ablation experiments on OLS model with same features as in Figure \ref{fig:sem}, reporting Multiple R$^2$ and $p$-value of the F-statistic. Left column: performance when only features of that group are used, right: when all features except that group are used. Significance levels: ***: $p$<0.0001, **: $p$<0.001, *: $p$<0.01, .: $p$<0.05.}
\label{tab:ablation}
\begin{tabular}{lrlrl}
  \toprule
  \textbf{Feature group} & \textbf{Group Only} &  & \textbf{Group Excluded} & \\
  \midrule
   Full model          & 0.658 & *** & 0.658 & *** \\
   Economic Inequality & 0.440 & *** & 0.632 & **  \\
   Demographic         & 0.492 & *** & 0.626 & *** \\
   Political Leaning   & 0.370 & *** & 0.639 & **  \\
   Gun Culture         & 0.453 & *** & 0.597 & *   \\
   Health              & 0.163 & .   & 0.655 & *** \\
   Twitter (all)       & 0.331 & *** & 0.638 & *** \\
  \bottomrule
\end{tabular}
\end{table*}

\subsection{Model 2: Individual March Attendance}


Next, we turn to examining the predictors of political engagement on the individual scale, in particular, predicting whether an individual declares one's attendance to the March for Our Lives on March 24, 2018.
Table~\ref{tab:userclassification} shows the performance of the three classifiers -- Support Vector Machine, Logistic Regression, and Random Forest -- when using all features, estimated via 5-fold cross-validation. 
The best performance in terms of F1 score is achieved by Logistic Regression at F1$= 0.807 (\pm0.022)$, mainly by having the best recall of the three ($0.794 (\pm0.042)$).
Hence, we use Logistic Regression model for the rest of the analyses.


\begin{table}[t]
\centering
\caption{Performance of classification of whether a user has gone to the march by 5-fold cross validation, $n=1206$.}
\label{tab:userclassification}
\begin{tabular}{lrrrr}
  \toprule
 & \textbf{Accuracy} & \textbf{Precision} & \textbf{Recall} & \textbf{F1} \\
 \midrule
Support Vector Machine & 0.801 ($\pm$0.021)  & 0.829 ($\pm$0.023)  & 0.781 ($\pm$0.038)  & 0.804 ($\pm$0.023) \\
Logistic Regression & 0.802 ($\pm$0.019)  & 0.822 ($\pm$0.021)  & 0.794 ($\pm$0.042)  & 0.807 ($\pm$0.022) \\
Random Forest & 0.793 ($\pm$0.006)  & 0.811 ($\pm$0.002)  & 0.788 ($\pm$0.013)  & 0.799 ($\pm$0.007) \\
\bottomrule
\end{tabular}
\end{table}

To deepen our understanding on which set of features have more predictive power, we conduct an ablation study of the proposed feature sets.  Table~\ref{tab:ablationstudy_marchprediction} shows the results of prediction models based on different set of features. First, when focusing on individual feature sets, we find that `Twitter Content (C)' feature set alone achieves a high prediction performance with F1$= 0.795 (\pm0.030)$. Then, the following three feature sets: Twitter Behavior (B), LIWC (L), and Demographics (D) marginally improve the model performance (0.011, 0.002, 0.010, respectively) when added to the model. The other features sets, such as Socio-Economics (S), Health (H), Politics - Gun Culture (P), and Twitter Network (N), in fact, degrades the model performances. The model with the four feature sets (CBDL) shows the best prediction performance with F1$= 0.818 (\pm0.020)$, which is higher than the model with all features. Next, we use the best model (CBDL) to investigate which features are most important in predicting individual political engagement.

\begin{table}[t]
\centering
\caption{Ablation study of the proposed feature sets based on Logistic Regression model for predicting  whether a user has gone to the march by 5-fold cross validation, $n=1206$.}
\label{tab:ablationstudy_marchprediction}
\begin{tabular}{lrrrr}
  \toprule
 \textbf{Feature sets} & \textbf{Accuracy} & \textbf{Precision} & \textbf{Recall} & \textbf{F1} \\
 \midrule
    Twitter Network (N) & 0.569 ($\pm$0.011)  & 0.633 ($\pm$0.015)  & 0.422 ($\pm$0.047)  & 0.504 ($\pm$0.032) \\
    Politics - Gun Culture (P) & 0.585 ($\pm$0.022)  & 0.616 ($\pm$0.020)  & 0.545 ($\pm$0.045)  & 0.578 ($\pm$0.032) \\
    Health (H) & 0.599 ($\pm$0.031)  & 0.629 ($\pm$0.032)  & 0.567 ($\pm$0.049)  & 0.596 ($\pm$0.038) \\
    Socio-Economics (S) & 0.604 ($\pm$0.037)  & 0.636 ($\pm$0.036)  & 0.569 ($\pm$0.065)  & 0.599 ($\pm$0.048) \\
    Demographics (D) & 0.596 ($\pm$0.018)  & 0.607 ($\pm$0.018)  & 0.651 ($\pm$0.040)  & 0.627 ($\pm$0.020) \\
    LIWC (L) & 0.677 ($\pm$0.018)  & 0.754 ($\pm$0.006)  & 0.569 ($\pm$0.050)  & 0.647 ($\pm$0.033) \\
    Twitter Behavior (B) & 0.704 ($\pm$0.017)  & 0.712 ($\pm$0.016)  & 0.729 ($\pm$0.028)  & 0.720 ($\pm$0.018) \\
    \bf Twitter Content (C) & 0.791 ($\pm$0.027)  & 0.816 ($\pm$0.017)  & 0.775 ($\pm$0.041)  & \bf 0.795 ($\pm$0.030) \\
\midrule
    C & 0.791 ($\pm$0.027)  & 0.816 ($\pm$0.017)  & 0.775 ($\pm$0.041)  & 0.795 ($\pm$0.030) \\
    CB & 0.803 ($\pm$0.008)  & 0.828 ($\pm$0.008)  & 0.786 ($\pm$0.026)  & 0.806 ($\pm$0.011) \\
    CBL & 0.806 ($\pm$0.015)  & 0.835 ($\pm$0.014)  & 0.784 ($\pm$0.036)  & 0.808 ($\pm$0.019) \\
    \bf CBLD & 0.813 ($\pm$0.020)  & 0.837 ($\pm$0.024)  & 0.800 ($\pm$0.029)  & \bf 0.818 ($\pm$0.020) \\
    CBLDS & 0.801 ($\pm$0.019)  & 0.826 ($\pm$0.019)  & 0.786 ($\pm$0.034)  & 0.805 ($\pm$0.022) \\
    CBLDSH & 0.806 ($\pm$0.020)  & 0.826 ($\pm$0.021)  & 0.797 ($\pm$0.034)  & 0.811 ($\pm$0.021) \\
    CBLDSHP & 0.803 ($\pm$0.018)  & 0.824 ($\pm$0.023)  & 0.794 ($\pm$0.038)  & 0.808 ($\pm$0.020) \\
    CBLDSHPN & 0.802 ($\pm$0.019)  & 0.822 ($\pm$0.021)  & 0.794 ($\pm$0.042)  & 0.807 ($\pm$0.022) \\
\bottomrule
\end{tabular}
\end{table}

\begin{table*}[t]
\centering
\caption{Important features of the best model for predicting individual march attendance (Logistic Regression with the four feature sets--Twitter Content, Twitter Behavior, LIWC, and Demographics (CBLD)). Top 20 most important features for predicting users who attended March For Our Lives and those not.}
\label{tab:march_prediction_important_features}
\begin{tabular}{p{0.47\textwidth} p{0.47\textwidth}}
  \toprule
  \textbf{Attended (+)} & \textbf{Did not attend (-)} \\
  \midrule
    liwc\_time, con\_hashtag\_entropy, liwc\_i, twt\_guntweetcnt, liwc\_we, liwc\_male, liwc\_hear, liwc\_relativ, liwc\_conj, liwc\_anger, liwc\_tentat, twt\_userfollowerscount, liwc\_posemo, liwc\_work, liwc\_achiev, twt\_accage, liwc\_space, liwc\_discrep, liwc\_article, liwc\_affiliation
    &
    liwc\_death, liwc\_motion, liwc\_insight, liwc\_shehe, liwc\_leisure, liwc\_family, liwc\_money, liwc\_assent, con\_rt\_entropy, liwc\_swear, liwc\_cause, liwc\_they, liwc\_adverb, twt\_engtweetcnt, liwc\_cogproc, liwc\_risk, liwc\_ingest, liwc\_differ, con\_voca, con\_voca\_2
    \\
  \bottomrule
\end{tabular}
\end{table*}

Examining the most important features based on the coefficients of variables in the best model (CBLD)
listed in Table~\ref{tab:march_prediction_important_features}, we find the LIWC categories to be among some of the most useful ones.
In particular, those who declared their attendance to the protest used more words concerning time, personal pronouns ``I'' and ``we'', expressed both anger and positive emotion in their prior tweets. 
They also used a greater variety of hashtags (hashtag entropy, con\_hashtag\_entropy) and had more tweets about the gun issue (twt\_guntweetcnt), which is well-aligned with the result in the previous literature, which finds that a large volumes of tweeting has a strong predictive power for forecasting state-level offline protest events~\cite{ertugrul2019activism}. 
Those who did not go to the march are more likely to tweet words associated with insight (about thinking or believing), and motion, as well as swear words and those of sad emotion.
Further, they are more likely to use ``he'', ``she'', and impersonal pronouns (``it'', ``everyone'', and ``what'').
They are also more likely to retweet many different sources (retweet entropy) and have more tweets in general. 
Note that, although demographic variables were present in the data, they were not as powerful predictors as these content-based ones.

To get the external validity of linguistic characteristics we found, we compare our results with other LIWC-driven studies in various social context.
First, it is well known that the usage of pronouns is different across social groups. 
For example, women and young people tend to use more first-person singular pronouns~\cite{graesser2014coh}.
The observation that our march-attending users also show higher usage of first-person singular pronouns (liwc\_i) is well aligned with a survey showing that 70 percent of participants were women~\cite{fisher2019american}.

We conduct a further analysis to see whether we indeed have more females among the march attenders in our data. In doing so, for those 1,206 users used in building the individual march attendance prediction model, we infer their gender using their self-reported first names in their Twitter profiles. 
Such approach has been popularly used to infer gender of social media users~\cite{amislove2011@icwsm,an2015whom}. 
In particular, we use a `Name gender classifier' that infers the gender based on a given first name~\cite{liu2013s}, which is a widely used approach in social media analytics~\cite{culotta2015predicting,hovy2015demographic}.
Out of 631 users who attended the march, 227 (35.97\%), 289 (48.80\%), 115 (18.23\%) are detected as male, female, and unknown, respectively. On the other hand, out of 575 users who did not attend the march, 230 (40\%), 182 (31.65\%), 163 (28.35\%) are detected as male, female, and unknown, respectively. 
First, we find that those who attended the march have more proportion of users with `inferred gender' when compared with those who did not attend the march. In other words, those who did not attend the march are less likely to have complete profiles, potentially avoiding to express their identities. Second, we observe that those who attended the march have higher proportion of female users than those who did not attend the march. After excluding those unknown users whose gender are not identified, we find that 56\% (44.17\%) of those who attended (did not attend) the march are females.
Finally, attempts have been made at automatically classifying age of Twitter users \cite{wang2019demographic}, however insufficient performance precludes analysis similar to gender.

Also, as March for Our Lives is a student-led movement from the start, it is reasonable to assume that young people actively engaged in online discussion and the march. 
Second, we compare our results with previous work from the perspectives of formality in discourse.
Formality entails that the focus of discourse is to be ``precise, coherent, articulate, and convincing to an educated audience''~\cite{graesser2014coh}.
By contrast, informal discourse has been characterized by a higher usage of swear words, internet slang, assents, and fillers.
Although not all four linguistic markers appear in our experiment, we find two of them (liwc\_swear, liwc\_assents) to be statically significantly more frequent in the vocabulary of non-attending users. 
This result indicates that users who attended the march show linguistic patterns that are closer to formal discourse, which aims to convince an educated audience, as opposed to non-attending users who are closer to informal discourse. 
Third, cognitive complexity has been modeled through usage of conjunctions, because they connect multiple thoughts and thus build coherent narratives~\cite{graesser2004coh}.
We find that march-attending users use more conjunctions (liwc\_conj) than non-attending users, thus indicating their online discussions show a richness of reasoning by differentiation or integration of thoughts through conjunctions. 
Finally, it is somewhat surprising to observe that march-attending users show more anger (liwc\_anger) and higher positive emotions (liwc\_posemo).
Both categories of words, however, have been reported as indicators of protests;  the anger is a good predictor of state-level Charlottesville and Ferguson protests~\cite{ertugrul2019activism}, and positive emotion could inspire collective action~\cite{goodwin2006emotions}.

The results show that users' online activity on Twitter, and language they use while participating in gun-related debates in particular, can be meaningful cues in understanding their offline behaviors.

\section{Discussion}

This work contributes to a rich multidisciplinary research line on the interplay between the use of social media and political engagement. 
Previous studies about the gun debate on social media are often descriptive analyses of content such as images~\cite{stefanone2015image} and text~\cite{wang2016machine}. 
Others move beyond description and correlate the opinions expressed on Twitter with those determined from polls; they find a moderate correlation at 0.51 Pearson $r$, despite using a low-recall (high-precision) hashtag matching approach~\cite{benton2016after}. 
The state-level model we present here puts into context the importance of Twitter as a factor in political engagement in the gun rights/control issue. 
We find that some of the strongest factors are socioeconomic (especially housing problems and income inequality ratio), but we find that political leaning as expressed on Twitter~\textemdash~more precisely the relative size of the two sides in the network~\textemdash~also has a substantial relationship with the laws (in fact, more than the percentage of state residents who voted for Republican in the previous election). 
Although the SEM structure represents our causal assumptions about the relationship among the variables, and we find supporting evidence for our model in the data, we cannot be certain that the model accurately represents reality, and we certainly do not make any causal claims about the findings.
Nevertheless, our findings suggest that there is a useful signal in Twitter data, as it pertains to concrete public policies and laws regarding gun control. 
Consequently, the outcomes of both state and individual models illustrate that political action can be captured on social media, beyond simple rhetoric or ``slacktivism''~\cite{vitak2011s}. 

Much of the research on political speech online deals with markers of political alignment~\cite{conover2011political,colleoni2014echo}, but the presence of these markers does not always lead to political engagement offline. 
An indirect relationship with political action (voting) was explored in early work on the prediction of election outcomes by using online political sentiment~\cite{Metaxas2011,gayo2012no,mejova2013gop,jungherr2012pirate}.
The fact that our automated classifier is able to predict attendance of a political march (albeit in a limited setting) suggests that social media provides signals not only of online ``activism'' but also of offline actions. 
Unlike previous studies, which either predict online political action of individual users (e.g., by looking at factors such as gender~\cite{bode2017closing} and personality traits~\cite{russo2016personality}), or predict offline political outcomes by looking at users collectively~\cite{Metaxas2011,gayo2012no,mejova2013gop,jungherr2012pirate,conover2011political,colleoni2014echo}, our study examines an individual social media user and their tweets indicating offline political action. 
Although we use aggregated census data for demographics here, it is possible to enrich the view of individuals with additional information, such as gender, age, and personality traits, all of which can be extracted from social media to a certain extent~\cite{vicente2019gender,chen2019joint,skowron2016fusing}.
Studying political action at an individual level could provide a detailed view to understand the major problems of political representation of under-served populations, political apathy, intimidation, and radicalization, which is not straightforward to be studied in a collective manner.
Of course, such research would need to be heavily scrutinized in order to maximally protect the privacy of individuals to be studied, for example, by pre-registering the methodology and receiving the approval of an Institutional Review Board (IRB).\footnote{As will be done by collaborators with Facebook who study the 2020 U.S. Election \url{https://about-fb-com.cdn.ampproject.org/c/s/about.fb.com/news/2020/08/research-impact-of-facebook-and-instagram-on-us-election/amp/}}

Considering the political issue of gun rights/control in particular, we observe the two sides of the argument to be well separated on the Twitter retweet network \textemdash a sign of the echo-chamber effect~\cite{conover2011political,garimella2018political,bakshy2015exposure,barbera2015tweeting}. Even though they do \emph{mention} prominent politicians from the other side, they do not interact with each other's content. 
This finding supports the observation that gun control is one of the most divisive political issues in the U.S.~\cite{nothing-divides}.
Recent research provides some support that social media may contribute to the worsening of polarization.
For instance, Republicans expressed substantially more conservative views after being exposed to opposing political views on social media (though no statistically significant effect was found for Democrats) \cite{bail2018exposure}.
Large-scale experiments on Facebook also showed that ``random variation in exposure to news on social media substantially affects the slant of news sites that individuals visit'' \cite{levy2021social} and that ``social media expose individuals to at least some ideologically crosscutting viewpoints'' but that ``individual choices more than algorithms limit exposure to attitude-challenging content'' \cite{bakshy2015exposure}.  
Still, these effects are not consistent among the platforms, as recently de-polarization was detected on the messaging service WhatsApp (although Twitter remained highly polarized) \cite{yarchi2021political}.
Further, because social media accounts for an estimated 4.2\% of total online consumption \cite{allen2020evaluating}, additional research on how other news and communication behaviors intersect with the polarization on the gun issue would allow for a broader understanding of its context.
For instance, local and national news consumption may impact beliefs around the prevalence of gun violence, which is a crucial concern for those leaning Democrat politically \cite{schaeffer2021key}.
With the decline of printed news media, the influence of social media on mainstream news is being documented around the globe (e.g. in China \cite{yan2020impact}, U.S. \cite{su2019agenda}, and Kashmir \cite{noor2017citizen}), thus it is important to understand whether potentially divisive stances affect other sources of information and public fora.

Another exciting future research direction is to observe whether these sides remain static if mass shootings keep rocking the country.
Historical survey data shows swings in the public opinion around mass shootings, but the overall trend is toward fewer restrictions on gun ownership~\cite{roper2016shootings}; longitudinal studies will reveal whether these short- and long-term trends are reflected on the internet, as well as at the ballot box.
Although one may find the binarization of such a complex topic to be an over-simplification, we point the reader to the high inter-annotator agreement on the stances of the users in this data, suggesting that, at least on Twitter, the issue has few discussants in the ``grey area'' of the opinion spectrum (similar findings about the polarization of this platform have been found in other domains, such as vaccination \cite{cossard2020falling}).



Within the network, we show that denser, smaller communities of gun control Twitter users are associated with more restrictive state-wide laws. 
We also find a lower degree of assortativity to be beneficial to their cause, which indicates more star-like structures around highly connected nodes (who may be leaders or facilitators bringing the community together).
Both these findings which echo Granovetter's ``embeddedness'' argument~\cite{granovetter1985economic}, according to which the causes of (in our case, political) action are not entirely determined by the individual (considered as rational and self-interested), nor determined by social categories only (e.g., by class or race); rather the action is `embedded' in the social structure, which requires an investigation of the social networks of individuals to be fully understood.
One possible explanation is that hierarchical structures have lower interdependence between the peers.
Therefore, a ``pooled'' interdependence structure that leverages the center of the network is more efficient~\cite{thompson2003organizations}. 
These findings are in contrast with a recent study of English Wikipedia that finds the successful editors collaborate in a more assortative, egalitarian network~\cite{rychwalska2020quality}. 
Thus, we urge future research to examine to what extent social media enables explicit coordination between users, and whether this is reflected in the hierarchical structure of its networks.

In the user model, content features were the most predictive of declaring offline political action on Twitter, and especially those concerning self-reference, as well as emotion. 
Previous literature shows excessive emotion to be associated with partisanship, negative when threatened and positive when reassured~\cite{huddy2015expressive}. In addition, emotion can also be spurred by events going against one's moral stance, thus resulting in a ``moral outrage''~\cite{mullen2006exploring}. 
However, greater emotion has been also associated with political ``sophistication'' (attention to, interest in, and knowledge of the political sphere)~\cite{miller2011emotional}.
In an age of increased populist political movements~\cite{caramani2019national,hawkins2019contemporary,jungherr2021populist}, it is imperative to understand the relationship between heightened emotions and political engagement, and the direction of their causality. 
Social media may provide a valuable record of people's expressed emotional states that can contribute to further understanding the role of emotion in political activity.

This study has several notable limitations, common to social media studies in general, and those on ongoing political issues in particular.
Only those with access to and interest in social media are included in our dataset~\cite{olteanu2016social}, as well as those who choose to be vocal on the subject. 
Thus, not everyone who attended the march was captured in our dataset. 
In fact, there are several ``filters'' which were used to select the users for the attendance classification, including self-selection by the users to post on Twitter, post specifically on the topic of interest, and, of course, indicate that they are going to the march. 
Also, these users have shared their location in their profile, which adds to the likelihood that these individuals are comfortable with sharing personal information (note that this information was shared by the user generally, not in the context of gun control). 
The results of the classification exercise, then, is applicable largely to those who are vocal enough to be included in the data, and open enough to share their approximate location. 
The task would be much more difficult if we were to look for potential march attendees in a generic Twitter stream.
Further, although our classifiers achieve a substantial accuracy (96\% using network and 70\% using content features), the user stances on the issue can only be approximated by using the limited information we have.
Similarly, the limited number of geographical divisions (in this case, states), does not allow for the inclusion of too many variables, thus forcing a strict feature selection before the SEM can be constructed, which limits the hypothesis space.
The selection of English language may also mean an exclusion of minority groups within the U.S. who use other languages -- additional data collection (using keywords in target languages) and NLP pipelines would need to be created to capture them.
Finally, one of the major limitations of this work is the relatively short span of the data, which does not reflect the historical political activity that has resulted in the current political climate around the gun rights/control issue. 
A longitudinal study over several years would be able to capture a greater variety of political activity and opinions expressed, and possibly even changes in opinion over time.

\textbf{Implications for design.} 
What we found, however, is that online discussions can be done in strikingly different ways across the regions if they have different offline contexts. 
In the experiment of predicting the strictness of gun control laws at the state level, we observe that people's interaction patterns (i.e., retweet) in online discussions are strongly coupled with their offline context (i.e., gun law rating of the state they live in). In other words, offline context defines what kinds of online discussion one would expect and experience.
Typically, there is a consensus that online platforms serve as an open place where the public can discuss social issues without limitations of geographic distances. 
Our finding, however, is that online discussions on certain issues, which are strongly coupled with state-level policies (e.g., gun control), can be shaped by geographical factor or other offline context.
This observation allows us to explore the potential of offline context for better online discussions on social media platforms. 
Let us simplify the challenge and start from a familiar instance of offline context, location. 
Incorporating the location of users into online platforms is not a new concept. In several online communities that mainly involve offline meetings, locations have  been a crucial factor for user experience.
For example, Meetup personalizes user experience based on locations and enable users to discover local events. 
Such platform proves that, when location is appropriately considered as a design choice, online platforms can \emph{connect} online and offline and provide continuous experience to users. 
Beyond these location-based services, our findings imply that conventional social media platforms also can benefit from considering offline context into their designs. 
While social media platforms use locations in a limited way, such as trending topics for a specific region, the benefit of locations can be extended.
As social issues become more frequently discussed online, the context and tone around discussion can be quite different across the regions (i.e., states) even for the same issue (i.e., debate on guns).
Consequently, how online platforms (1) consider offline context that might lead to nuanced discussions, (2) make users aware of it, and (3) allow them not to mislead others' opinions or stance by avoiding wrong assumptions about offline context will be a crucial design choice to support better discussions online.

Moreover, it is interesting that predicting individual-level tweeting about their attendance of the march was highly accurate (though on a class-balanced sample, ``real-world'' performance may be much lower).
The prediction model, however, should be implemented with extra caution. 
As social media has become the place to empower social movements by grassroots activists in recent years, predicting who will or did attend offline protests from  online data could be misused in both prospective and retrospective ways. 
This risk can be managed by giving early warnings to users when needed so that they know the risk in advance, seek more secure channels, and so on.

\textbf{Privacy and ethics.} 
This study has been performed on publicly available tweets and user information, as Twitter API does not provide any data on private accounts or personal messages between users. 
The fact that, using this data, our automated classifier is able to predict whether users attend a political march and make it known through a public tweet, which suggests that social media provides signals not only of online activism but also of offline actions. 
It is therefore reasonable to assume that government or political actors are already using such techniques,\footnote{\url{https://www.bbc.com/news/blogs-china-blog-48552907}}$^{,}$\footnote{\url{https://www.reuters.com/article/us-iran-internet/irans-guards-increase-monitoring-of-social-media-state-tv-idUSKBN0LY1YC20150302}} so it is important both to raise awareness of the public about the risk of online activities that can be used to predict their offline political activities in the future, and to outline restrictions to large-scale data analysis by powerful entities. 
Finally, since current study does not interact with subjects, nor attempts to extract personally identifiable information and uses only public data, this study does not fall in the category of human participant research. Instead, the authors has followed the ethics guidelines outlined by their institutions. Furthermore, our study adheres to the Twitter Terms of Service for the publicly available data through their API.\footnote{\url{https://developer.twitter.com/en/developer-terms/agreement-and-policy}}

\appendix

\section{Features}\label{app:features}
We list all features used for Model 2. Individual March Attendance below.

\begin{longtable}{p{0.45\textwidth} p{0.45\textwidth}}
  \toprule
\textbf{Demographics} & \textbf{LIWC} \\
perc under 18 & Total function words (liwc\_function) \\
perc 65 and over & Total pronouns (liwc\_pronoun) \\
perc African American & Personal pronouns (liwc\_ppron) \\
perc Asian & 1st person singular (liwc\_i) \\
perc Hispanic & 1st person plural (liwc\_we) \\
perc non Hispanic white & 2nd person (liwc\_you) \\
perc rural & 3rd person singular (liwc\_shehe) \\
 & 3rd person plural  (liwc\_they) \\
\textbf{Socio-Economic} & Impersonal pronouns (liwc\_ipron)\\
high school graduation rate & Articles (liwc\_article) \\
perc some college & Prepositions (liwc\_prep) \\
perc unemployment & Auxiliary verbs (liwc\_auxverb) \\
income inequality ratio & Common Adverbs (liwc\_adverb)\\
perc uninsured & Conjunctions (liwc\_conj) \\
perc single parent households & Negations (liwc\_negate) \\
perc association rate & Common verbs  (liwc\_verb) \\
violent crime rate & Common adjectives (liwc\_adj) \\
perc severe housing problems & Comparisons (liwc\_compare) \\
median household income & Interrogatives (liwc\_interrog) \\
residential segregation black white & Numbers (liwc\_number)\\
homicide rate & Quantifiers (liwc\_quant) \\
 & Affective processes (liwc\_affect) \\
\textbf{Health} & Positive emotion (liwc\_posemo) \\
mentally unhealthy days & Negative emotion (liwc\_negemo) \\
perc adult smoking & Anxiety (liwc\_anx) \\
perc adult obesity & Anger (liwc\_anger) \\
perc excessive drinking & Sadness (liwc\_sad) \\
 & Social processes (liwc\_social) \\
\textbf{Politics \& Gun Culture} & Family (liwc\_family) \\
gun sales & Friends (liwc\_friend) \\
firearm fatalities rate & Female references (liwc\_female) \\
perc vote republican & Male references (liwc\_male) \\
\textbf{Twitter Network} & Cognitive processes (liwc\_cogproc) \\
indicator whether in GCC (net\_inGCC) &  Insight (liwc\_insight) \\
& Causation (liwc\_cause) \\
\textbf{Twitter Content} & Discrepancy (liwc\_discrep) \\ 
retweet count (con\_rt\_account) & Tentative (liwc\_tentat) \\
retweet entropy (con\_rt\_entropy) & Certainty (liwc\_certain) \\
hashtag count (con\_hashtag\_count) & Differentiation (liwc\_differ) \\
hashtag entropy (con\_hashtag\_entropy) &  Perceptual processes (liwc\_percept) \\
vocabulary count (con\_voca) & See (liwc\_see) \\
vocabulary entropy (con\_voca\_2) & Hear (liwc\_hear) \\
hate word rate (con\_hateword) & Feel (liwc\_feel) \\
avg sentiment polarity (con\_sentiment) & Biological processes (liwc\_bio \\
& Body (liwc\_body) \\
\textbf{Twitter Behavior} & Health (liwc\_health)\\
user followers count & Sexual (liwc\_sexual) \\
user friends count & Ingestion (liwc\_ingest) \\
follower/friend ratio & Drives  (liwc\_drives) \\
account age (twt\_accage) & Affiliation (liwc\_affiliation) \\
gun tweet rate (twt\_guntweetrate) & Achievement (liwc\_achiev) \\
all tweet rate (twt\_alltweetrate) & Power (liwc\_power) \\
gun tweet count (twt\_guntweetcnt) & Reward (liwc\_reward) \\
English tweet count (twt\_engtweetcnt) & Risk (liwc\_risk) \\
 & Past focus (liwc\_focuspast) \\
 & Present focus (liwc\_focuspresent) \\
 & Future focus  (liwc\_focusfuture) \\
 & Relativity (liwc\_relativ) \\
 & Motion (liwc\_motion) \\
 & Space (liwc\_space) \\
 & Time (liwc\_time) \\
 & Work (liwc\_work) \\
 & Leisure (liwc\_leisure) \\
 & Home (liwc\_home) \\
 & Money (liwc\_money) \\
 & Religion (liwc\_relig) \\
 & Death (liwc\_death) \\
 & Informal language (liwc\_informal) \\
 & Swear words (liwc\_swear) \\
 & Netspeak (liwc\_netspeak) \\
 & Assent (liwc\_assent) \\
 & Nonfluencies (liwc\_nonflu) \\
 & Fillers (liwc\_filler) \\
\bottomrule
\end{longtable}


\bibliographystyle{ACM-Reference-Format}
\bibliography{guns}


\end{document}